\documentclass[twocolumn,aps,nofootinbib]{revtex4-1}
\usepackage{amsfonts, amssymb}
\usepackage{graphicx, epsfig,bm}
\usepackage{color}

\usepackage{amsbsy}
\usepackage{amsfonts}
\usepackage{threeparttable}
\usepackage{amsmath}
\usepackage{amssymb}
\usepackage{url}

\newcommand{\be}{\begin{equation}}
\newcommand{\ee}{\end{equation}}
\newcommand{\bea}{\begin{eqnarray}}
\newcommand{\eea}{\end{eqnarray}}

\def\fun#1#2{\lower3.6pt\vbox{\baselineskip0pt\lineskip.9pt
        \ialign{$\mathsurround=0pt#1\hfill##\hfil$\crcr#2\crcr\sim\crcr}}}

\newcommand{\neff}{\ensuremath{N_\mathrm{eff}}}
\newcommand{\sumnu}{\ensuremath{\sum{m_{\nu}}}}

\newcommand{\nrun}{\ensuremath{dn_s/d\ln k}}

\newcommand\lsim{\mathrel{\rlap{\lower4pt\hbox{\hskip1pt$\sim$}}
    \raise1pt\hbox{$<$}}}
\newcommand\gsim{\mathrel{\rlap{\lower4pt\hbox{\hskip1pt$\sim$}}
    \raise1pt\hbox{$>$}}}

\def\dslash{\not{\hbox{\kern-2pt $\partial$}}}
\def\Dslash{\not{\hbox{\kern-4pt $D$}}}
\def\Oslash{\not{\hbox{\kern-4pt $O$}}}
\def\Qslash{\not{\hbox{\kern-4pt $Q$}}}
\def\pslash{\not{\hbox{\kern-2.3pt $p$}}}
\def\kslash{\not{\hbox{\kern-2.3pt $k$}}}
\def\qslash{\not{\hbox{\kern-2.3pt $q$}}}

 \newtoks\slashfraction
 \slashfraction={.13}
 \def\slash#1{\setbox0\hbox{$ #1 $}
 \setbox0\hbox to \the\slashfraction\wd0{\hss \box0}/\box0 }

\def\ee{\end{equation}}
\def\be{\begin{equation}}

\begin{document}
\setlength{\unitlength}{1mm}
\title{Are Light Sterile Neutrinos Preferred or Disfavored by Cosmology?}
\author{Shahab Joudaki, Kevork N. Abazajian, Manoj Kaplinghat}
\affiliation{Center for Cosmology, Dept.~of Physics \& Astronomy, University of California, Irvine, CA 92697}

\date{\today}

\begin{abstract}
We find that the viability of a cosmological model that incorporates 2 sterile neutrinos with masses around 1 eV each, as favored by global neutrino oscillation analyses including short baseline results, is significantly dependent on the choice of datasets included in the analysis and the ability to control the systematic uncertainties associated with these datasets. Our analysis includes a variety of cosmological probes 
including the cosmic microwave background (WMAP7+SPT), Hubble constant (HST), galaxy power spectrum (SDSS-DR7), and supernova distances (SDSS and Union2 compilations).
In the joint observational analysis, our sterile neutrino model is equally favored as a $\Lambda$CDM model
when using the MLCS light curve fitter for the supernova measurements, and strongly disfavored by the data at $\Delta\chi^2_{\rm eff} \approx 18$ when using the SALT2 fitter.
When excluding the supernova measurements, the sterile neutrino model is disfavored by the other datasets at $\Delta\chi^2_{\rm eff} \approx 12$, and at best becomes mildly disfavored at $\Delta\chi^2_{\rm eff} \approx 3$ when allowing for curvature, evolving dark energy, additional relativistic species, running of the spectral index, and freedom in the primordial helium abundance. 
No single additional parameter accounts for most of this effect. Therefore, if laboratory experiments continue to favor a scenario with roughly eV mass sterile neutrinos, and if this becomes decisively disfavored by cosmology, then a more exotic cosmological model than explored here may become necessary.
\end{abstract}
\bigskip

\maketitle

\section{Introduction}
\label{introlabel2v}

The standard models of particle physics and cosmology do not yet fully describe the neutrino sector, with open questions related to the mass-generation mechanism of the neutrinos, any sterile neutrino partners of the active neutrinos, and their potential relation to the number of relativistic degrees of freedom inferred from cosmology.
In recent years, there has been some experimental evidence pointing towards the existence of additional light (effectively massless) degrees of freedom. 
In particular, a combined analysis of cosmic microwave background (CMB) data from WMAP7, baryon acoustic oscillation (BAO) distances from SDSS+2dF, and Hubble constant from HST yields a weak preference for additional light degrees of freedom ($\neff = 4.34 \pm 0.87$)~\cite{Komatsu:2010fb}. When moreover including small-scale CMB data from ACT or SPT, this preference mildly increases to the $2\sigma$ level ($\neff = 4.56 \pm 0.75$ with addition of ACT~\cite{dunkleyact} and $\neff = 3.86 \pm 0.42$ with addition of SPT~\cite{Keisler:2011aw}). 
These constraints on $\neff$ explicitly assume that the additional particles are massless, and have sparked further work~\cite{Joudaki:2012fx,Hamann:2010bk,Archidiacono:2011gq,Riess:2011yx,Smith:2011es,Hou:2011ec, Hamann:2007pi, Hamann:2011hu, GonzalezMorales:2011ty, Calabrese:2011hg,Smith:2011ab, Hamann:2011ge,Giusarma:2011ex,Giusarma:2011zq,Fischler:2010xz,deHolanda:2010am,Nakayama:2010vs, Burenin:2012uy, Ciuffoli:2012yd}.

In light of new predictions for the anti-neutrino flux from nuclear reactors, 
global short-baseline neutrino oscillation data now favor the existence of two sterile neutrinos with best-fit masses 
of $m_4 = 0.68~{\rm{eV}}$ and $m_5 = 0.94~{\rm{eV}}$, assuming massless active neutrinos~\cite{Kopp:2011qd} (also see~\cite{Giunti:2011gz,Donini:2012tt,Abazajian:2012ys,Conrad:2012qt}).
Instead of analyzing the data with the aim of estimating an upper bound to the mass of an additional thermalized neutrino species~\cite{Hamann:2010bk,Giusarma:2011zq,Giusarma:2011ex}, we take the existence of two sterile neutrinos with $m_4$ and $m_5$ as a prior assumption consistent with the short-baseline data. 
It is our aim to determine how a model with these two additional neutrino species fares compared to the case without them, when including all available and relevant cosmological data.

\begin{table}[t!]
\vspace{0.5em}
\begin{center}
\begin{tabular}{lc|c}
\hline\hline
Parameter & Symbol & Prior\\
\hline
Baryon density & $\Omega_{b}h^2$ & $0.005 \to 0.1$\\
Cold dark matter density & $\Omega_{c}h^2$ & $0.01 \to 0.99$\\
Angular size of sound horizon & $\theta_s$ & $0.5 \to 10$\\
Optical depth to reionization & $\tau$ & $0.01 \to 0.8$\\
Scalar spectral index & $n_{s}$ & $0.5 \to 1.5$\\
Amplitude of scalar spectrum & $\ln{(10^{10} A_{s})}$ & $2.7 \to 4$\\
\hline
Effective number of neutrinos & \neff &  $3.046 \to 10$\\
{\it~~~--~with sterile neutrinos} & \neff &  $5.046 \to 10$\\
Sum of neutrino masses & $\sum{m_{\nu}}~\rm{[eV]}$ &  $0$\\
{\it~~~--~with sterile neutrinos} & $\sum{m_{\nu}}~\rm{[eV]}$ &  $1.62$\\
Constant dark energy EOS & $w$ & $-3 \to 0$\\
Running of the spectral index & ${dn_s \over d\ln k}$ &  $-0.4 \to 0.4$\\
Curvature of the universe & $\Omega_{k}$ &  $-0.4 \to 0.4$\\
Primordial helium abundance & $Y_p$ &  $0 \to  1$\\
\hline\hline
\end{tabular}
\caption{We impose uniform priors on the above cosmological parameters. 
In addition, we always consider the Poisson point source power $D_{3000}^{\rm{PS}}$, the clustered power $D_{3000}^{\rm{CL}}$, and the SZ power $D_{3000}^{\rm{SZ}}$ as nuisance parameters constrained by the CMB data~\cite{Keisler:2011aw}.
Moreover, we always derive $\sigma_8$, the amplitude of linear matter fluctuations on scales of $8~{\rm{Mpc}}/h$ at $z=0$.
We only vary a redshift-independent dark energy equation of state (EOS).
In this table, the first 6 parameters are defined as ``vanilla" parameters.
}
\vspace{-2.5em}
\label{table:priorsnu}
\end{center}
\end{table}

\begin{figure*}[!t]
\begin{center}
\vspace{-0.4em}
\includegraphics[bb=4.540570 10.481554 672.656604 470.818040,clip,scale=0.35]{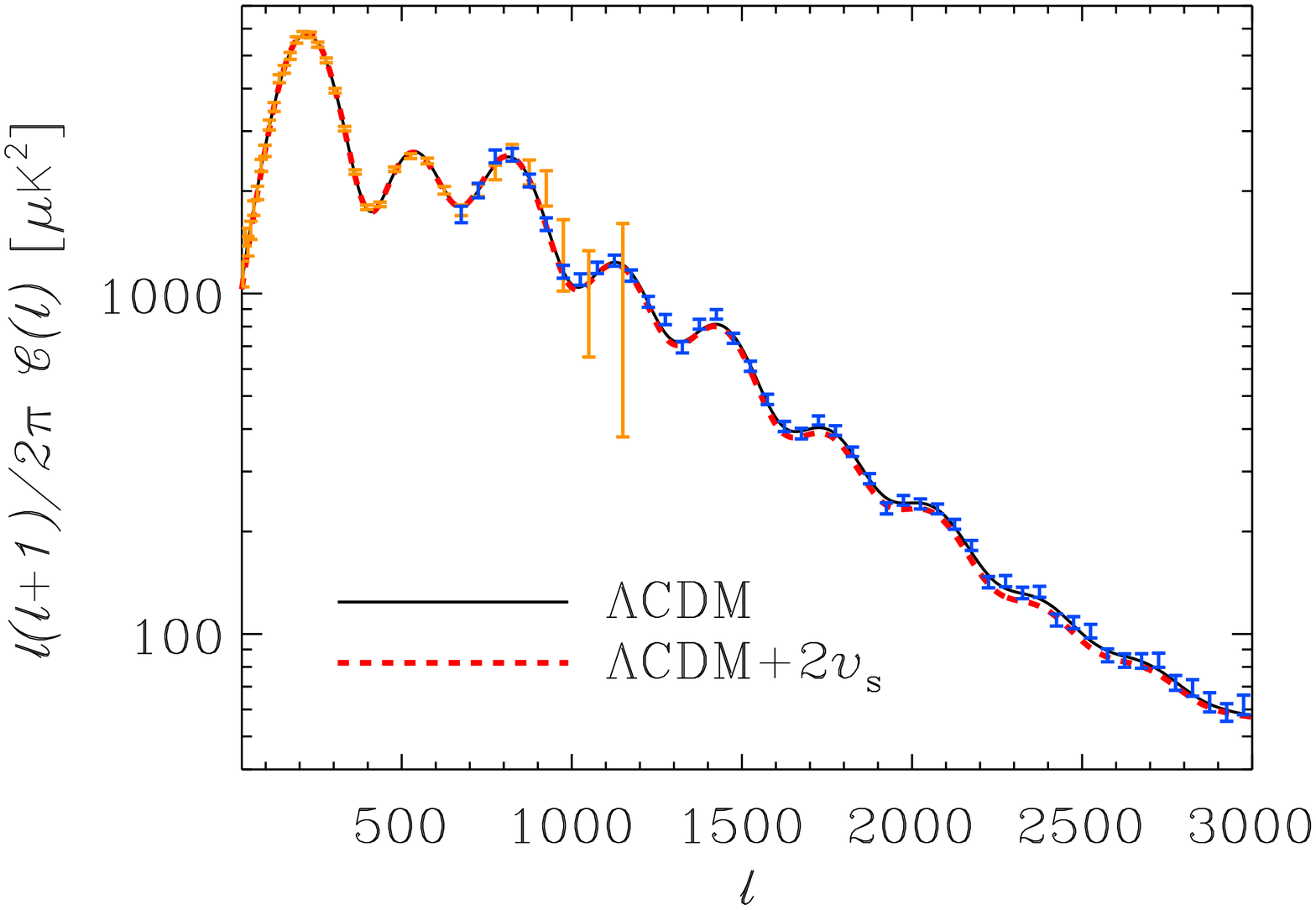}
\includegraphics[bb=1.138570 3.064570 668.518081 476.020110,clip,scale=0.35]{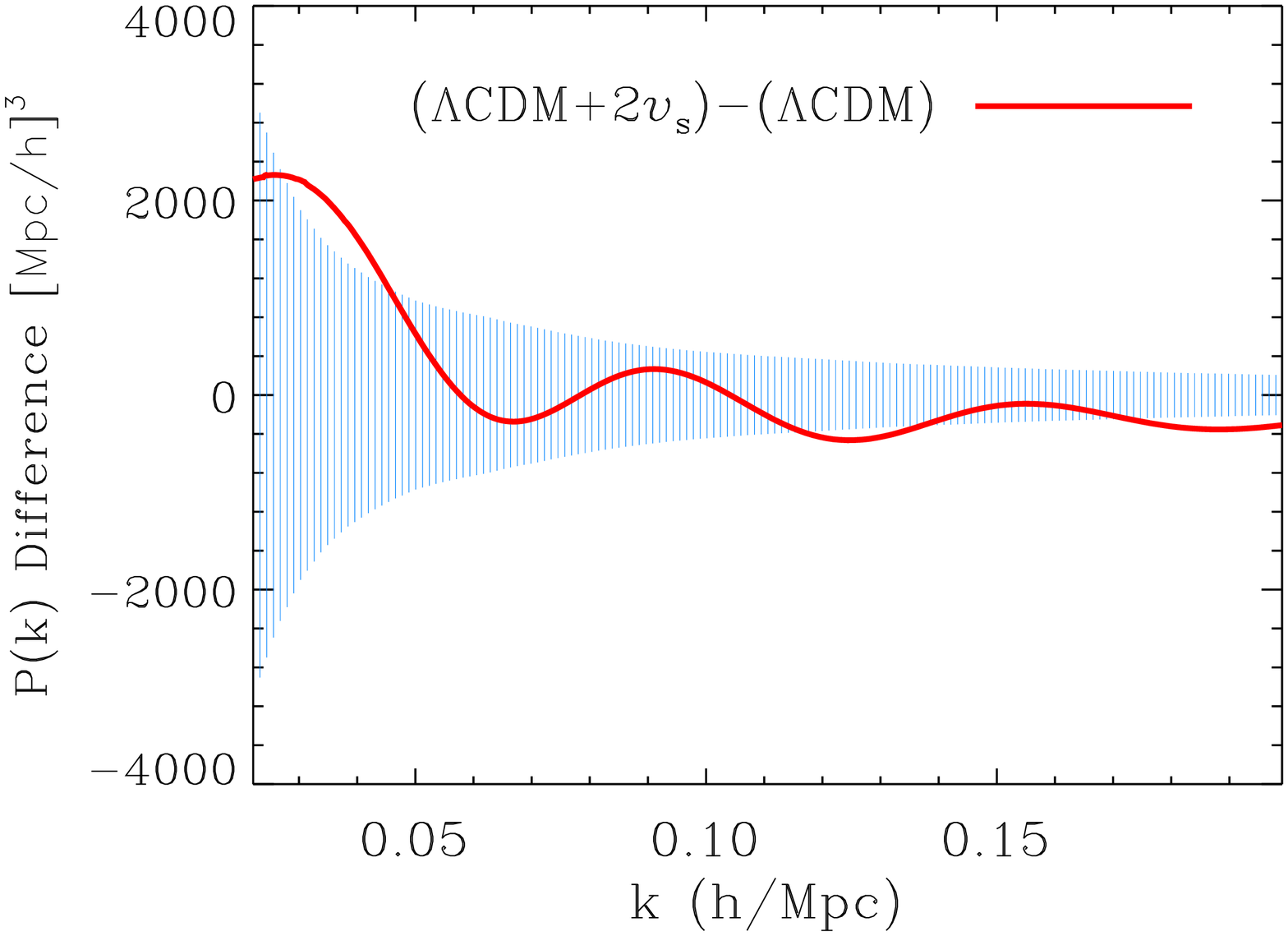}
\end{center}
\vspace{-1.9em}
\caption{
{\it Left}: CMB temperature power spectrum measurements with WMAP7 (orange) and SPT (blue). The $\Lambda$CDM model without sterile neutrinos is shown with the solid (black) line, and the $\Lambda$CDM model with 2 sterile neutrinos is shown in dashed (red).
{\it Right}: Assuming the $\Lambda$CDM model is centered on the DR7 data, with error bars given by the shaded band (in blue), we show the absolute difference with our sterile neutrino model in solid (red).
}
\label{fig:cmbpknu}
\end{figure*}

We examine the impact of the two sterile neutrinos on other cosmological parameters in the vanilla $\Lambda$CDM model, such as the matter density, amplitude of linear matter fluctuations on $8~{\rm{Mpc}}/h$ scales, and spectral index. We also explore the impact of extensions of a cosmological model with sterile neutrinos, including nonzero curvature, evolving dark energy, running of the spectral index, and primordial helium abundance. 
Throughout this paper, we will assume that the two sterile neutrinos are thermally populated as seems reasonable given the large mixing angles of the sterile neutrinos~\cite{DiBari:2001ua,Abazajian:2002bj}. If this is not the case, then the differences between a model with two sterile neutrinos and one without them will be smaller (cf.~Refs.~\cite{Abazajian:2004aj,Hannestad:2012ky,Mirizzi:2012we}).

The cosmological influence of sterile neutrinos includes an increase in the effective number of neutrinos to $\neff = 5.046$ and the sum of neutrino masses to $\sumnu = 1.62~{\rm{eV}}$ assuming full thermalization.
As discussed in Ref.~\cite{Joudaki:2012fx}, the effective number of neutrinos is mainly correlated with the matter density and spectral index in a vanilla $\Lambda$CDM model. 
In extended cosmological models, correlations also exist with the helium abundance, dark energy equation of state, and running of the spectral index.
Meanwhile, the sum of neutrino masses is mainly correlated with the matter density and Hubble constant in a vanilla $\Lambda$CDM model, along with the dark energy equation of state and curvature density in extended parameter spaces~\cite{Joudaki:2012fx}.

The radiation content of the universe can be constrained from big bang nucleosynthesis (BBN) through its effect on the expansion rate~\cite{Kneller:2004jz,Simha:2008zj,Steigman:2007xt}.
Given the standard BBN consistency relation between the set of parameters $\{Y_p, \neff, \Omega_b h^2\}$~\cite{Simha:2008zj}, 
the inclusion of 2 additional neutrinos boosts the primordial helium abundance by $\Delta Y_p = 0.024$ when the baryon density is kept fixed. 
Thus, $Y_p \approx 0.27$ in standard cosmological analyses when enforcing this consistency relation.
Primordial helium abundance estimations from observations of metal poor extragalactic H~II regions suffer from significant systematic uncertainties (e.g.~see~\cite{Peimbert:2007vm,Izotov:2007ed,Izotov:2010ca,Aver:2010wq,Aver:2010wd,Aver:2011bw}). 
An extensive analysis that attempts to account for these systematic uncertainties gives $Y_p = 0.2534 \pm 0.0083$~\cite{Aver:2011bw}, which is consistent with the cosmological estimate at 95\% CL (assuming 5 light neutrinos).
This agreement could be tightened by lowering $Y_p$ from cosmology, achieved via mechanisms such as incomplete thermalization, presence of a non-zero chemical potential, or post-BBN production of the sterile neutrinos from the decay of a heavy particle species (e.g.~see~\cite{Hamann:2011ge}).

We describe our analysis method in Section~2. 
In Section~3, we provide constraints on a $\Lambda$CDM model with three massless active neutrinos and two massive sterile neutrinos, and determine how well this model fits cosmological data relative to a model without sterile neutrinos.
We further explore to what extent the tension between the two models could be ameliorated by an extension of parameter space including evolving dark energy, universal curvature, running of the spectral index, additional relativistic species, and freedom in the primordial helium abundance (all parameters defined in Table~\ref{table:priorsnu}).
Section~4 concludes with a discussion of our findings. 

\section{Methodology}
\label{secmeth}

We employed a modified version of CosmoMC~\cite{Lewis:2002ah,cosmomclink} in performing Markov Chain Monte Carlo (MCMC) analyses of parameter spaces with sterile neutrinos, using CMB data from WMAP7~\cite{Komatsu:2010fb} and SPT~\cite{Keisler:2011aw}, luminous red galaxy power spectrum measurements from SDSS DR7~\cite{Reid:2009xm}, the Hubble constant from HST~\cite{Riess:2011yx}, and SN distances from either the Union2 compilation~\cite{Amanullah:2010vv} or the SDSS compilation~\cite{Kessler:2009ys}. 
We generally impose a cutoff in the galaxy power spectrum measurements at $k = 0.1~h/{\rm Mpc}$ because of insufficient understanding of the matter power spectrum on nonlinear scales when including baryons, massive neutrinos, and dark energy~\cite{Joudaki:2011nw, Smith, Rudd, vanDaalen:2011xb,McDonald,Joudaki, Saito:2008bp,Brandbyge:2008rv,Wong:2008ws,Saito:2009ah}. For the same reasons, we do not include the small-scale power spectrum from Lyman-$\alpha$ forest data. 

The Union2 compilation consists of 557 SNe, which includes large samples from SCP, SNLS, ESSENCE, HST, and older datasets~\cite{Amanullah:2010vv}, while the SDSS compilation consists of 288 SNe from SDSS, SNLS, ESSENCE, HST, and a set of low-redshift SNe~\cite{Kessler:2009ys}.
For the Union2 compilation, we considered the SALT2 light curve fitter~\cite{Guy:2007dv}, while for the SDSS compilation, we considered both the MLCS~\cite{Jha:2006fm, Kessler:2009ys} and SALT2 fitters.
The two fitting methods estimate cosmological parameters in different ways, make different assumptions about the nature of color variations in type Ia SNe, and employ different training procedures that determine the spectral and light-curve templates~\cite{Kessler:2009ys}.

More specifically, MLCS returns the value and uncertainty of the distance modulus for each SN (marginalizing over model parameters), the set of which are then included in the cosmological analysis, while SALT2 determines the distance moduli along with cosmological and SN parameters in a global fit to all of the light curves~\cite{Kessler:2009ys}.
Further, MLCS assumes that excess color variation is entirely due to extinction by dust, 
and therefore imposes a positivity prior on the extinction, such that it is effectively zero for SNe with apparent colors that are bluer than the templates~\cite{Kessler:2009ys,Guy:2010bc}. 
Meanwhile, apparently blue SNe are assigned negative colors in SALT2, such that the respective luminosities and distance moduli are larger than those from MLCS~\cite{Kessler:2009ys}.

Moreover, while MLCS trains on a sample of nearby SNe, and extends the measured relationship between light-curve shape and color to higher redshift SNe, the training procedure for SALT2 uses a combination of both low and high redshift data~\cite{Kessler:2009ys,Foley:2010mm}. 
Given the systematic discrepancies in rest-frame U-band between the nearby and higher redshift samples, much of the difference in the estimated best-fit cosmology between MLCS and SALT2 may further be traced to the respective U-band models determined in the training~\cite{Kessler:2009ys}.
At present, there seems to exist no consensus on which light curve fitter is the most accurate~(e.g.~\cite{rjfoley,Kessler:2009ys}).

\begin{figure}[!t]
\vspace{-0.3cm}
\epsfxsize=3.7in
\centerline{\epsfbox{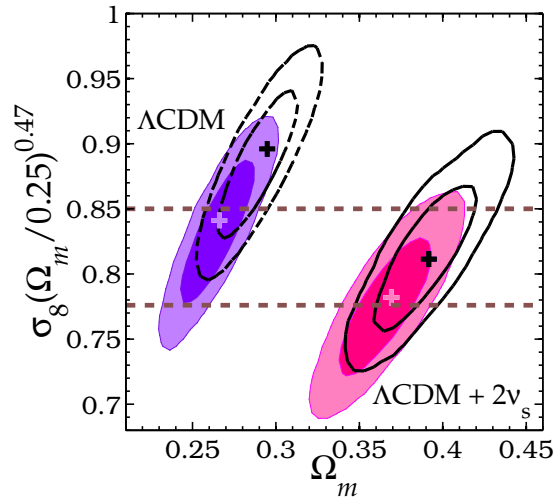}}
\vspace{-0.3cm}
\caption{
Joint two-dimensional marginalized constraints on $\sigma_8 (\Omega_m/0.25)^{0.47}$ against $\Omega_m$ from combining the measurements of 
WMAP+SPT+$P(k)$+HST+SNe. The purple and pink shaded confidence regions (inner 68\%, outer 95\%) are obtained using 
SNe from the Union2 compilation (SALT2), while the solid and dashed transparent ellipses are obtained using SNe from the SDSS compilation (MLCS).
The overlapping ellipses preferring a lower matter density ({\it left}) are for the $\Lambda$CDM model without sterile neutrinos, while the overlapping ellipses 
preferring a larger matter density ({\it right}) are for the $\Lambda$CDM model with sterile neutrinos.
The horizontal dashed lines (in brown) denote the 68\% confidence interval about the mean from the local ($0.025 < z < 0.25$) galaxy cluster abundance measurement of Vikhlinin~et~al.~(2009)~\cite{Vikhlinin:2008ym}.
}
\label{fig:sig8nu}
\end{figure}

\begin{table*}[t!]\footnotesize
\begin{center}
\caption{Constraints on Cosmological Parameters using SPT+WMAP+$P(k)$+$H_0$. In some of the columns, we further add SNe from either the Union2 or SDSS compilations. The foreground priors on the SZ, poisson point sources, and clustering point sources are encapsulated in ``FG."}
\begin{tabular}{lccc|cc|ccccccccc|}
\hline \hline
 & & $\Lambda$CDM & $\Lambda$CDM & $\Lambda$CDM & $\Lambda$CDM & $\Lambda$CDM & $\Lambda$CDM\\
 & & & +2$\nu_s$ & +SNe$_{\rm Union2}$ & +2$\nu_s$+SNe$_{\rm Union2}$ & +SNe$_{\rm SDSS}$ & +2$\nu_s$+SNe$_{\rm SDSS}$\\
\hline \hline
Primary & $100 \Omega_b h^2$ & $2.242 \pm 0.039$ & $2.296 \pm 0.040$ & $2.241 \pm 0.039$ & $2.308 \pm 0.040$ & $2.225 \pm 0.038$ & $2.293 \pm 0.038$ \cr
& $100\Omega_{\rm dm} h^2$ & $11.15 \pm 0.32$  & $16.49 \pm 0.41$ & $11.18 \pm 0.29$  & $16.09 \pm 0.36$ & $11.63 \pm 0.29$ & $16.51 \pm 0.36$ \cr
& $10^4\theta_s$ & $104.15 \pm 0.15$ & $103.86 \pm 0.15$ & $104.15 \pm 0.15$  & $103.90 \pm 0.15$ & $104.10 \pm 0.15$ & $103.85 \pm 0.15$ \cr
& $\tau$ & $0.086 \pm 0.014$ & $0.089 \pm 0.014$ & $0.087 \pm 0.014$ & $0.092 \pm 0.015$ &  $0.082 \pm 0.013$ & $0.089 \pm 0.014$ \cr
& $100n_s$ & $96.73 \pm 0.95$ & $98.32 \pm 0.97$ & $96.69 \pm 0.94$ & $98.81 \pm 0.95$ & $95.98 \pm 0.94$ & $98.27 \pm 0.94$ \cr
& $\ln{(10^{10} A_{s})}$ & $3.187 \pm 0.035$ & $3.214 \pm 0.036$ & $3.189 \pm 0.035$ & $3.195 \pm 0.036$ & $3.216 \pm 0.035$ & $3.216 \pm 0.035$ \cr
\hline
Derived & $H_0$ & $71.4 \pm 1.4$ & $69.6 \pm 1.3$ & $71.2 \pm 1.3$ & $71.0 \pm 1.2$ & $69.2 \pm 1.2$ & $69.5 \pm 1.2$ \cr
 & $\sigma_8 (\Omega_m/0.25)^{0.47}$ & $0.829 \pm 0.039$ & $0.813 \pm 0.039$ & $0.833 \pm 0.035$ & $0.776 \pm 0.034$ & $0.886 \pm 0.036$ & $0.816 \pm 0.036$ \cr
\hline
$\chi^2_{\rm eff}$ & CMB & 7512.4 & 7517.2 & 7511.7 & 7517.9 & 7513.2 & 7516.7 \cr
& $P(k)$ & 23.9 & 28.9 & 24.5 & 30.2 & 23.2 & 28.7 \cr
& $H_0$ & 1.5 & 2.9 & 1.2 & 1.5 & 4.6 & 3.5 \cr
& SNe & --- & --- &  530.8 & 536.0 & 245.9 & 237.9 \cr
& FG & 0.1 & 0.6 & 0.4 & 0.7 & 0.1 & 0.7 \cr
& Total & 7537.9 & 7549.5 & 8068.5 & 8086.2 & 7787.0 & 7787.4 \cr
DIC & Total & 7554.1 & 7566.1 & 8085.2 & 8103.0 & 7803.4 & 7804.1 \cr
\hline
$\Delta\chi^2_{\rm eff}$ & Total & --- & 11.6 & --- & 17.7 & --- & 0.4 \cr
$\Delta$DIC & Total & --- & 12.0 & --- & 17.8 & --- & 0.7 \cr
\hline
\hline
\end{tabular}
\begin{tablenotes}
\item Mean of the posterior distribution of cosmological parameters along with the symmetric 68\% confidence interval about the mean. The three active neutrinos are taken to be massless in all models. We also consider adding 2 sterile neutrinos (denoted as ``2$\nu_s$") of masses $m_{\nu_{s1}} = 0.68$ and $m_{\nu_{s2}} = 0.94$, such that the sum of neutrino masses is $\sumnu = 1.62~{\rm eV}$. We fix the primordial helium mass fraction $Y_p = 0.2478$. The Deviance Information Criterion is defined as ${\rm{DIC}} = 2 \overline{\chi^2_{\rm eff}(\theta)} - {\chi^2_{\rm eff}(\hat{\theta})}$, where $\theta$ is the vector of varied parameters, the bar denotes the mean over the posterior distribution, and hat denotes the maximum likelihood point. For the SDSS SNe, we have used the MLCS light curve fitter. The corresponding total $\Delta\chi^2_{\rm eff}$ and $\Delta$DIC values when using the SALT2 fitter are $\Delta\chi^2_{\rm eff} = 20.1$ and ${\Delta}{\rm DIC} = 19.4$. For the Union2 SNe, we always use the SALT2 fitter. In the rows with $\chi^2_{\rm eff}=-2\ln {\cal L}_{\rm max}$ values listed for individual probes, the values are computed at the maximum likelihood estimate (MLE) for the joint analysis including all probes. If each probe is analyzed separately, the MLE will be different and the corresponding $\Delta\chi^2_{\rm eff}$ values will be smaller.
\end{tablenotes}
\label{table:chival}
\end{center}
\end{table*}

All parameters are defined in Table~\ref{table:priorsnu}.
The power spectra of the CMB temperature and E-mode polarization were obtained from a modified version of the Boltzmann code CAMB~\cite{LCL,camblink}. 
We used the Gelman and Rubin $R$ statistic~\cite{gelmanrubin} to determine the convergence of our chains, where $R$ is defined as the variance of chain means divided by the mean of chain variances. 
In stopping the runs, we generally required the conservative limit $(R - 1) < 10^{-2}$, 
and checked that further exploration of the tails does not change our results.

In our baseline $\Lambda$CDM model, we include 3 massless neutrinos. 
We also consider an expanded $\Lambda$CDM model that contains 2 sterile neutrinos in addition to the 3 active neutrinos of the baseline model. 
The sterile neutrino masses are given by the mass splittings with the lightest neutrino mass: 
$m_{4} = 0.68~{\rm{eV}}$ and $m_{5} = 0.94~{\rm{eV}}$~\cite{Kopp:2011qd}. 
Beyond the 3 massless active species and 2 massive sterile species, additional contributions to $\neff$ are assumed massless.

For the primordial fraction of baryonic mass in helium, there are three reasonable priors we can explore: 1) fixing $Y_p$ to a constant, 2) allowing $Y_p$ to vary as a free parameter, and 3) determining $Y_p$ as a function of $\{\neff, \Omega_b h^2\}$ in a manner consistent with BBN (e.g.~see Eqn~1 in Ref.~\cite{Joudaki:2012fx}).
We show results when fixing the the primordial helium abundance to the SPT preferred value of $Y_p = 0.2478$~\cite{Keisler:2011aw}. We have checked that our results do not significantly vary when forcing $Y_p$ to preserve the standard BBN consistency relation instead.
As part of our analysis of extended parameter spaces, we also consider cases with the helium abundance as an unknown parameter to be determined by the data.

We define the running of the spectral index $\nrun$ through the dimensionless power spectrum of primordial curvature perturbations:
\begin{equation}
\Delta^{2}_{R}(k) = \Delta^{2}_{R}(k_0)\left(\frac{k}{k_0}\right)^{n_s-1+\frac{1}{2}\ln(k/k_0) dn_s/d\ln k},
\end{equation}
where the pivot scale $k_0 = 0.002/{\rm{Mpc}}$. Due to the large correlation between $n_s$ and $\nrun$ at this scale, we consistently quote our values for $n_s$ at a scale $k_0 = 0.015/{\rm{Mpc}}$, where the tilt and running are less correlated, such that 
$n_s(k_0=0.015/{\rm{Mpc}}$) = $n_s(k_0=0.002/{\rm{Mpc}}$$) + \ln(0.015/0.002)dn_s/d\ln k$~\cite{Cortes:2007ak}. An example of the remaining correlation between the spectral index and its running is shown in Ref.~\cite{Joudaki:2012fx}.

We define $\chi^2_{\rm eff} = -2 \ln {\mathcal{L}}_{{\rm max}}$, where ${\mathcal{L}}_{\rm max}$ is the maximum likelihood of the data given the model.
The ratio of maximum likelihoods given two separate models is then ${{\mathcal{L}}_{\rm max,2}/{\mathcal{L}}_{\rm max,1}} = \exp({-\Delta\chi^2_{\rm eff}/2})$. For the case where $\Delta\chi^2_{\rm eff} > 0$, we interpret model 2 to be associated with a lower probability of drawing the data at the maximum likelihood point than model 1, by a factor given by $\exp({-\Delta\chi^2_{\rm eff}/2})$. For reference, a value of $\Delta\chi^2_{\rm eff} = 10$ corresponds to odds of 1 in 148, which we take as strong preference for model 1 as compared to model 2. 

\begin{figure*}[!t]
\begin{center}
\vspace{-0.4em}
\includegraphics[bb=112.066098 232.605063 461.807916 544.013913,clip,scale=0.53]{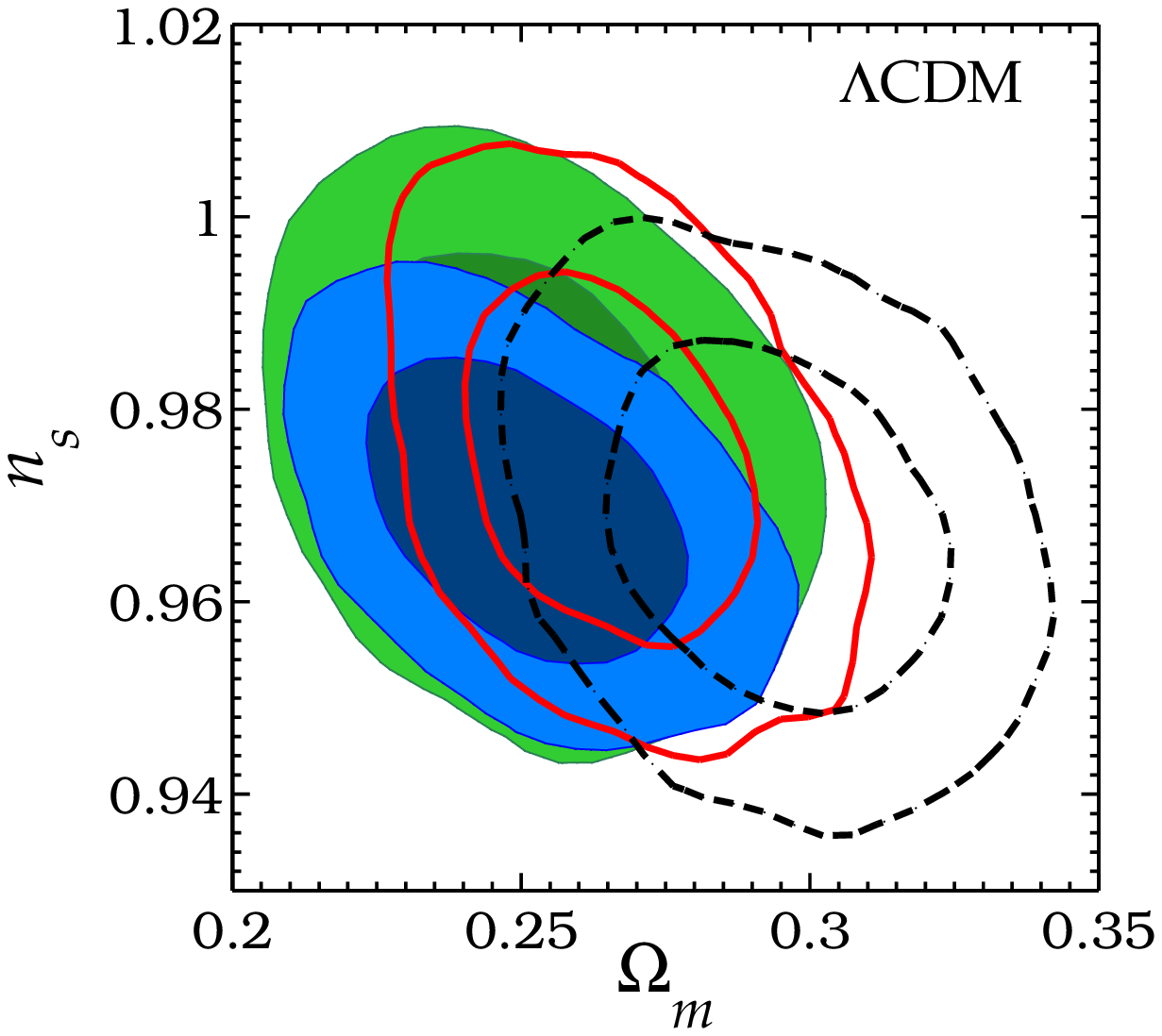}
\includegraphics[bb=112.066098 232.605063 445.414979 544.013913,clip,scale=0.53]{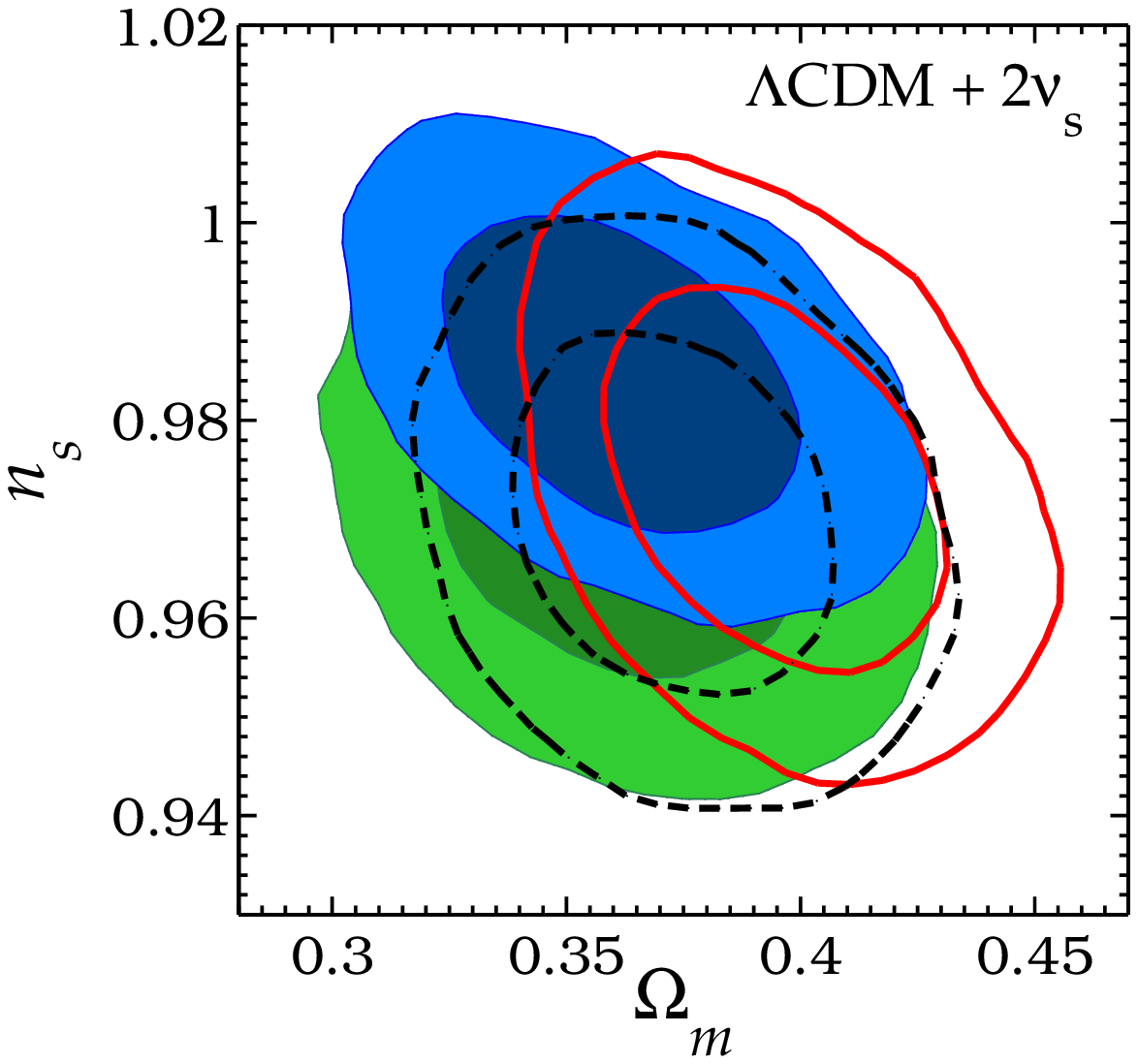}
\end{center}
\vspace{-1.9em}
\caption{
Joint two-dimensional marginalized constraints on the spectral index $n_s$ and matter density $\Omega_m$ (inner 68\%, outer 95\%).
The green shaded ellipses are for WMAP+HST, blue shaded ellipses are for WMAP+SPT+HST, the solid transparent ellipses (in red) are for WMAP+$P(k)$+HST, and the dashed transparent ellipses (in black) are for WMAP+HST+SNe, where
the SNe are from the SDSS compilation (MLCS). The panel to the {\it left} assumes a $\Lambda$CDM model without sterile neutrinos, while the panel to the {\it right} includes two sterile neutrinos.
}
\label{fig:consistnu}
\end{figure*}

We also consider the Deviance Information Criterion (DIC)~\cite{spiegelhalter},
given by ${\rm{DIC}} = {\chi^2_{\rm eff}(\hat{\theta})} + 2C_b$, where $C_b = \overline{\chi^2_{\rm eff}(\theta)} - {\chi^2_{\rm eff}(\hat{\theta})}$ is the so-called ``Bayesian complexity," such that
$\theta$ is the vector of varied parameters, the bar denotes the mean over the posterior distribution, and hat denotes the maximum likelihood point~\cite{Trotta:2008qt}. 
The Bayesian complexity can be thought of as the effective number of unconstrained parameters, such that it penalizes more complex models with more parameters, independently of how well the models fit the data~\cite{bridlelewis}.
If the Bayesian complexity of two models is the same, the difference in DIC between the models matches their difference in $\chi^2_{\rm eff}$ values.
We take a difference beyond 10 in DIC values between two models to constitute a strong preference for one model as compared to the second model, with the more preferred model being the one with the smaller DIC value.

\section{Results}
\label{secres}

We now explore the cosmological constraints on our sterile neutrino models, and the relative goodness of fit with respect to models without sterile neutrinos. In Sec.~\ref{onesec}, we vary the parameters of a vanilla model defined in Table~\ref{table:priorsnu}, while we consider an extended parameter space in Sec.~\ref{twosec}.

\subsection{Vanilla plus 2$\nu_{\rm s}$ Models}
\label{onesec}

In Table~\ref{table:chival}, we show the constraints on two separate $\Lambda$CDM models for three distinct supernova cases:~1)~without SNe, 2)~with Union2 SNe (SALT2 fitter), and 3)~with SDSS SNe (MLCS fitter).~The model denoted ``$\Lambda$CDM" consists of the 6 vanilla parameters in Table~\ref{table:priorsnu} and does not contain sterile neutrinos, while the model denoted ``$\Lambda{\rm{CDM}} + 2\nu_s$" consists of the same vanilla parameters but now contains two sterile neutrinos of fixed masses $m_4 = 0.68~{\rm{eV}}$ and $m_5 = 0.94~{\rm{eV}}$ (as discussed in Sec.~\ref{secmeth}). 
We define $\Delta\chi^2_{\rm eff}$ as being the difference in $\chi^2_{\rm eff}$ between the sterile neutrino model ($\Lambda{\rm{CDM}} + 2\nu_s$) with the null model ($\Lambda{\rm{CDM}}$).

When excluding SN data, we find that the model with sterile neutrinos is disfavored at $\Delta\chi^2_{\rm eff} = 11.6$, which implies a factor of 330 larger odds for the null model to draw the data than the sterile neutrino model assuming the maximum likelihood parameters. 
Moreover, $\Delta{\rm DIC} = 12.0$.
Since $\Delta{\rm DIC} \simeq \Delta\chi^2_{\rm eff}$, this tells us that the two models have essentially the same Bayesian complexity, and both statistical measures (DIC and $\chi^2_{\rm eff}$) strongly favor the null model over the one with two massive sterile neutrinos.

Allowing for SN data, the corresponding results are $\Delta\chi^2_{\rm eff} = 17.7$ and $\Delta{\rm DIC} = 17.8$ for the Union2 compilation (SALT2 fitter), while $\Delta\chi^2_{\rm eff} = 0.4$ and $\Delta{\rm DIC} = 0.7$ for the SDSS compilation (MLCS fitter).
When using SDSS SN data with the SALT2 light curve fitter, the corresponding results are $\Delta\chi^2_{\rm eff} = 20.1$ and $\Delta{\rm DIC} = 19.4$. 
In other words, $[{\Delta\chi^2_{\rm eff}{\rm (Union2_{SALT2})} \approx \Delta\chi^2_{\rm eff}{\rm (SDSS_{SALT2})} \approx 20}] > [{\Delta\chi^2_{\rm eff}{\rm (SDSS_{MLCS})} \approx 0}]$.
{\it{Thus, the choice of light curve fitter has a decisive impact on the statistical viability of the sterile neutrino model.}}
The two fitters are also associated with slight differences in the inferred matter density and Hubble constant, with larger values of the former and smaller values of the latter being associated with the MLCS fitter.
These discrepancies may ultimately be traced back to the use of color priors and differences between the fitters in the rest-frame U-band region~\cite{Kessler:2009ys,Foley:2010mm,Guy:2010bc}.

\begin{table*}[t!]\footnotesize
\begin{center}
\caption{Constraints on Cosmological Parameters using SPT+WMAP+$P(k)$+$H_0$. In some of the columns, we further add SNe from the Union2 compilation adopting the SALT2 fitter. Note that the MLCS fitter does not disfavor the addition of 2 light sterile neutrinos and those results are not shown here since the main motivation is to gauge the effect of additional cosmological parameters.}
\begin{tabular}{lccc|ccccccccccc|}
\hline \hline
 & & $\Lambda$CDM & $\Lambda$CDM & $\Lambda$CDM & $\Lambda$CDM\\
 & & & +2$\nu_s$ & +SNe$_{\rm Union2}$ & +2$\nu_s$+SNe$_{\rm Union2}$\\
\hline \hline
Primary & $100 \Omega_b h^2$ & $2.235 \pm 0.073$ & $2.232 \pm 0.067$ & $2.235 \pm 0.076$ & $2.208 \pm 0.069$ \cr
& $100\Omega_{\rm dm} h^2$ & $13.7 \pm 1.2$  & $19.8 \pm 1.4$ & $12.88 \pm 0.93$  & $18.6 \pm 1.0$ \cr
& $10^4\theta_s$ & $103.63 \pm 0.37$ & $103.03 \pm 0.28$ & $103.84 \pm 0.35$  & $103.17 \pm 0.29$ \cr
& $\tau$ & $1.002 \pm 0.019$ & $0.090 \pm 0.016$ & $0.093 \pm 0.016$ & $0.084 \pm 0.014$ \cr
& $100n_s$ & $97.5 \pm 2.7$ & $97.9 \pm 2.1$ & $97.3 \pm 2.3$ & $96.2 \pm 1.7$ \cr
& $\ln{(10^{10} A_{s})}$ & $3.105 \pm 0.068$ & $3.183 \pm 0.064$ & $3.156 \pm 0.045$ & $3.237 \pm 0.044$ \cr
\hline
Extended & $\Omega_k$ & $0.023 \pm 0.021$ & $0.019 \pm 0.022$ & $0.003 \pm 0.010$ & $-0.004 \pm 0.012$ \cr
 & $w$ & $-0.76 \pm 0.20$ & $-0.80 \pm 0.32$ & $-0.999 \pm 0.099$ & $-1.32 \pm 0.16$ \cr
& $\neff$ & $4.22 \pm 0.74$ & $6.83 \pm 0.97$ & $3.77 \pm 0.50$ & $6.02 \pm 0.67$ \cr
& $\nrun$ & $-0.048 \pm 0.036$ & $-0.029 \pm 0.027$ & $-0.027 \pm 0.033$ & $-0.019 \pm 0.027$ \cr
& $Y_p$ & $0.165 \pm 0.084$ & $0.086 \pm 0.061$ & $0.214 \pm 0.078$ & $0.113 \pm 0.067$ \cr
\hline
Derived & $H_0$ & $72.6 \pm 2.4$ & $72.8 \pm 2.5$ & $73.6 \pm 2.2$ & $74.8 \pm 2.1$ \cr
 & $\sigma_8 (\Omega_m/0.25)^{0.47}$ & $0.841 \pm 0.076$ & $0.828 \pm 0.070$ & $0.898 \pm 0.058$ & $0.875 \pm 0.050$ \cr
\hline
$\chi^2_{\rm eff}$ & Total & 7533.7 & 7540.9 & 8065.0 & 8074.9 \cr
DIC & Total & 7557.7 & 7565.7 & 8089.8 & 8097.1 \cr
\hline
\hline
\end{tabular}
\begin{tablenotes}
\item The models are the same as in Table~\ref{table:chival}, but here we consider an extended parameter space. We do not fix the primordial helium mass fraction, but instead allow it to vary as a free parameter given the condition $Y_p \geq 0$. We have imposed $\neff \geq 3.046$ for the cases without sterile neutrinos, and $\neff \geq 5.046$ for the cases with sterile neutrinos. For the cases without SNe, $\Delta\chi^2_{\rm eff} = 7.2$ and $\Delta{\rm DIC} = 8.0$. For the cases with SNe from the Union2 compilation, $\Delta\chi^2_{\rm eff} = 9.9$ and $\Delta{\rm DIC} = 7.3$. Further, excluding SN data and comparing the sterile neutrino case in an expanded space to the case without sterile neutrinos in a minimal space (Table~\ref{table:chival}), we find $\Delta\chi^2_{\rm eff} = 3.0$ and $\Delta{\rm DIC} = 10.8$. Including Union2 SNe and comparing the sterile neutrino case in an expanded space to the case without sterile neutrinos in a minimal space, we find $\Delta\chi^2_{\rm eff} = 6.4$ and $\Delta{\rm DIC} = 11.9$. When adding either $\neff$ or $\Omega_k$ as a single additional parameter to either of the cases with sterile neutrinos, there is a roughly $2\sigma$ preference above the null value. Including Union2 SNe and adding $w$ as a single additional parameter to the sterile neutrino case, there is a $2.7\sigma$ preference for $w < -1$. This preference is also visible in the analysis with the full extended parameter space shown in this table.
\end{tablenotes}
\label{table:chivalext}
\end{center}
\end{table*}

We have also explored to what extent the different results between the SALT2 and MLCS fitters are affected by the choice of SN datasets included in the analysis. To this end, we focused on the SDSS compilation, which is composed of 33 nearby SNe (${0.02 < z < 0.12}$), 103 SDSS SNe (${0.05 < z < 0.42}$), 56 ESSENCE SNe (${0.16 < z < 0.70}$), 62 SNLS SNe (${0.25 < z < 1.01}$), and 34 HST SNe (${0.22 < z < 1.55}$)~\cite{Kessler:2009ys}. We find that $\Delta\chi^2_{\rm eff} \approx 12$ for both light curve fitters when only one these SN datasets is included in the analysis. In other words, when only a single SN dataset is included, we find a much smaller difference in $\Delta\chi^2_{\rm eff}$ between the light curve fitters, and a much smaller difference in $\Delta\chi^2_{\rm eff}$ to the case where we do not include any SNe in the analysis. A large difference in $\Delta\chi^2_{\rm eff}$ between the two light curve fitters is manifested when combining SN data that cover a large range of redshifts, which minimally includes combinations such as the nearby+HST datasets, or the SDSS+ESSENCE+SNLS datasets.

We note that our results using the Union2 dataset are in agreement with those in Ref.~\cite{Kristiansen:2011mp}, which used SNe from the Union2 compilation~\cite{joskri}, extended the galaxy power spectrum measurements out to $k = 0.2~h/{\rm Mpc}$, and excluded small-scale CMB data. When extending the power spectrum measurements from $k_{\rm max} = 0.1~h/{\rm Mpc}$ out to $k_{\rm max} = 0.2~h/{\rm Mpc}$, $\Delta\chi^2_{\rm eff}$ increases by about 5 for all of the different cases including SN data.

In Fig.~\ref{fig:cmbpknu}, we show the CMB temperature and galaxy power spectra for a $\Lambda$CDM model without sterile neutrinos and one with 2 sterile neutrinos. While the influence of additional neutrinos is a systematic suppression in both spectra, the figures show that this level of suppression largely lies within the error bars of present data. In other words, the figures show that the sterile neutrino model provides a good fit to the data, albeit slightly worse than the null model. 
In Table~\ref{table:chival}, we directly show from which probes the largest differences in $\chi^2_{\rm eff}$ arise for our two models. 
For the case without SNe, $\Delta\chi^2_{\rm eff}$ receives an equal contribution of about 5 from each of the CMB and galaxy probes. For the case with SNe from the Union2 compilation, $\Delta\chi^2_{\rm eff}$ receives an equal contribution of about 6 from each of the CMB and galaxy probes, and roughly 5 from the SNe measurements. 
Hence, no single probe manages to decisively disfavor the sterile neutrino model.

For the case with SNe from the SDSS compilation, $\Delta\chi^2_{\rm eff}$ receives a contribution of 5.5 from $P(k)$, 3.5 from the CMB, but then a negative contribution of 8.0 from the SN measurements (MLCS). Thus, the main difference between our Union2 and SDSS supernova cases (with the different light curve fitters) is that the former disfavors sterile neutrinos, while the latter prefers sterile neutrinos. 
We note that the individual $\chi^2_{\rm eff}$ values in Table~\ref{table:chival} are those associated with the maximum likelihood point of the joint analysis of all considered probes. 
When each probe is analyzed separately, the best-fit $\Delta\chi^2_{\rm eff}$ values are less pessimistic.

In Table~\ref{table:chival}, we further show the constraints on a range of cosmological parameters. In particular, we find that the sterile neutrino model prefers a larger matter density and lower value of $\sigma_8$, while 
preserving the constraint on $\sigma_8 (\Omega_m/0.25)^{0.47}$ near the 0.8-mark, in agreement with the galaxy cluster abundance measurement of Vikhlinin~et~al.~(2009)~\cite{Vikhlinin:2008ym}.
In Fig.~\ref{fig:sig8nu}, we show error ellipses in the plane of $\sigma_8 (\Omega_m/0.25)^{0.47}$ and $\Omega_m$ for the case of WMAP+SPT+$P(k)$+HST+SNe. 
Remarkably, when using SNe from the SDSS compilation, a much larger matter density is allowed to constitute the energy content of our universe.


In our MCMC analyses, we minimize the total $\chi^2_{\rm eff}$, which is the sum of the individual $\chi^2_{\rm eff}$ values of equally weighted probes. 
However, given the different systematics,  the motivation for weighting the CMB and large-scale structure probes equally is not clear. In this regard, in Fig.~\ref{fig:consistnu}, we show error ellipses for $n_s$ against $\Omega_m$ given different sets of probes: 1) WMAP+HST, 2) WMAP+HST+SPT, 3) WMAP+HST+$P(k)$, and 4) WMAP+HST+SNe.
We find that these different combinations of probes constrain a portion of parameter space in agreement with each other, even for the case of sterile neutrinos.
In other words, the preferred parameter space is not driven by a single probe, but consistently preferred by all probes.

We have shown that the possibility of a cosmological model that incorporates 2 sterile neutrinos with roughly eV masses is significantly dependent on the choice of datasets included in the analysis and the ability to control the systematic uncertainties associated with these datasets. 
Concretely, the choice of light curve fitter in the analysis of SN data dictates whether the sterile neutrino model is favored in a combined analysis of CMB, galaxy power spectrum, Hubble constant, and SN data. However, the sterile neutrino model is disfavored if we exclude SN measurements from the analysis. In an attempt to reconcile with the laboratory preference for the massive sterile neutrino scenario, we proceed to explore if this is an indication of new physics beyond the standard cosmological model.

\subsection{Extended Cosmological Parameter Space \\2$\nu_{\rm s}$ Models}
\label{twosec}

As discussed in Sec.~\ref{onesec}, in a combined analysis of datasets that include the CMB, galaxy power spectrum, Hubble constant, and SN distances, we obtain different results with regard to the viability of sterile neutrinos depending on the choice of SN light curve fitter. 
In an analysis without SNe, sterile neutrinos 
are disfavored at $\Delta\chi^2_{\rm eff} = 11.6$ and $\Delta{\rm DIC} = 12.0$. We examined to what extent this tension could be alleviated in an expanded cosmological parameter space. 
As summarized in Table~\ref{table:chivalext}, we allow for variations in the universal curvature density, constant dark energy equation of state, running of the spectral index, additional relativistic species, and primordial helium abundance.

Excluding SN data, adding a single additional parameter to the sterile neutrino case does not decrease $\chi^2_{\rm eff}$ by a significant amount. For the case of $w$ or $\Omega_k$, we find a decrease in $\chi^2_{\rm eff}$ by about 2, while for $\nrun$, $\neff$, or $Y_p$ we find a decrease in $\chi^2_{\rm eff}$ by about 1. For the joint addition of all five of these parameters in the model with sterile neutrinos, we find that $\chi^2_{\rm eff}$ decreases by 8.6, such that $\Delta\chi^2_{\rm eff} = 3.0$ with respect to the $\Lambda$CDM model without sterile neutrinos and no additional parameters. However, due to a nonzero Bayesian complexity~(see Sec.~\ref{secmeth}), we still find a large $\Delta{\rm DIC} = 11.5$. 
Hence, including additional parameters to the sterile neutrino model decreases $\Delta\chi^2_{\rm eff}$ to a reasonable level, 
but the fact that the additional parameters are not well constrained is reflected in the pessimistic DIC estimates.

Accounting for the same parameter extension ($w, \Omega_k, \neff, \nrun, Y_p$) when adding SN distances from the Union2 compilation (i.e.~considering WMAP+SPT+$P(k)$+HST+SNe), we find a decrease in $\Delta\chi^2_{\rm eff} = 6.4$ (down from 17.7) and $\Delta{\rm DIC} = 11.9$ (down from 17.8).
When replacing the SNe from Union2 with those from SDSS-SALT2, we find $\Delta\chi^2_{\rm eff} = 7.4$ (down from 20.1) and $\Delta{\rm DIC} = 13.8$ (down from 19.4).
Hence, when accounting for SNe with the SALT2 fitter, an extended parameter space is unable to allow for our 2 massive sterile neutrinos.

Given the differences between HST and SDSS on the best estimate of the Hubble constant~\cite{Riess:2011yx,Sanchez:2012sg}, we considered removing the HST prior on $H_0$ from our analysis.
We found that excluding the $H_0$ prior does not significantly change our constraints, mainly because the HST prior only manages to boost the best estimate of $H_0$ by about 1~km/s/Mpc with respect to the value favored by the CMB and large-scale structure data. 
For instance, considering WMAP+SPT+$P(k)$+SNe, where the SNe are from the Union2 compilation, the $H_0$ constraint lies around 70~km/s/Mpc without an HST prior, and 71~km/s/Mpc when we impose the prior with central value around 74 km/s/Mpc. The latter is because the data constrains $H_0$ more strongly than the prior (such that the error bars on $H_0$ without the prior are about 1.4~km/s/Mpc, to be compared with the prior of 2.4~km/s/Mpc).
This line of reasoning works even when excluding SN data.
For the particular case WMAP+SPT+$P(k)$, we find $\Delta\chi^2_{\rm eff} = 9.6$ (down from 11.6) when not including the HST prior.

We also considered replacing the $P(k)$ measurements (with cutoff at $k = 0.1~h/{\rm Mpc}$) with two BAO distances from SDSS+2dFGRS~\cite{Percival:2009xn}. 
Considering the combination WMAP+SPT+HST+BAO, $\Delta\chi^2_{\rm eff} = 9.5$ (down from 11.6).
Hence, our results are robust to the choice of using the power spectrum or BAO distances.
Moreover, to obtain a better sense of the quoted $\chi^2_{\rm eff}$ values, we note that a universe with $w = -1/3$ is disfavored by $\Delta\chi^2_{\rm eff} = 96$ as compared to a universe with $w = -1$ (considering WMAP+SPT+$P(k)$+HST).
For a less extreme case, a universe with $w = -0.8$ is disfavored by $\Delta\chi^2_{\rm eff} = 9.4$ (with respect to $w = -1$). 
These $\Delta\chi^2_{\rm eff}$ values significantly increase when further including SN data.
Hence, our sterile neutrino model is disfavored at roughly the same level as a dark energy model with $w = -0.8$ (when not including SN data).

When forcing the 2 sterile neutrinos to be massless, $\Delta\chi^2_{\rm eff} = 5.9$ and $\Delta{\rm DIC} = 5.5$ (when not including SN data). 
Hence, roughly half of the degradation in $\chi^2_{\rm eff}$ and ${\rm DIC}$ could be captured by increasing $\neff$ by 2.
We note that adding two sterile neutrinos with a given total mass $m = m_4 + m_5$ is preferred to adding one sterile neutrino with mass $m$. For example, $\Delta\chi^2_{\rm eff}$ is lower by about 8 when $\{\neff = 5, m_{1,2,3} = 0, m_4 = 0.68~{\rm eV}, m_5 = 0.94~{\rm eV}\}$ as compared to $\{\neff = 4, m_{1,2,3} = 0, m_4 = 1.62~{\rm eV}\}$. 
However, for a given $\neff$, the data prefers the sum of neutrino masses to be distributed in the least number of neutrinos. For instance, given $\neff = 5$, we find that $\Delta\chi^2_{\rm eff}$ is lower by about 4 when $\{m_{1,2,3,4} = 0, m_5 = 1.62~{\rm eV}\}$ as compared to $\{m_{1,2,3} = 0, m_4 = 0.68~{\rm eV}, m_5 = 0.94~{\rm eV}\}$.

The 3+2 sterile neutrino model is preferred by cosmology as compared to a 3+1 model if the sum of neutrino masses is the same for the two models.
However, the 3+2 model is disfavored as compared to a 3+1 neutrino model with $m_4 = 1~{\rm eV}$, at the level of $\Delta\chi^2_{\rm eff} = 5.6$ when not including SN data, at the level of $\Delta\chi^2_{\rm eff} = 3.3$ when including SDSS-MLCS SN data, and at the level of $\Delta\chi^2_{\rm eff} = 9.1$ when including Union2-SALT2 SN data.

Perhaps more importantly, even when assuming the existence of two massive sterile neutrinos, we find a $2\sigma$ preference for an additional massless species. 
Thus, a model containing 3 sterile neutrinos (for example, see 3+3 models in Ref.~\cite{Conrad:2012qt}) is not necessarily ruled out by cosmology, especially if the sum of neutrino masses is not increased as compared to models with fewer number of sterile neutrinos.
However, if laboratory data converge on a $3+2$ or $3+3$ model with a larger sum of sterile neutrino masses than considered here, this model would have a larger difficulty to fit the cosmological data.
At about the 2$\sigma$ level, we also note that the extended parameter space model with two light sterile neutrinos shows a preference for super-acceleration (or $w<-1$)~\cite{Kaplinghat:2003vf}. In fact, this slight preference for $w<-1$ also persists in a model (with the two light sterile neutrinos) that is enlarged only by this one parameter ($w$).

To summarize, we have studied in detail the question of whether two sterile neutrinos with about eV mass  each is consistent or disfavored by the latest cosmological data.
While our sterile neutrino model fits each dataset well, in a combined analysis of the CMB, Hubble constant, and galaxy power spectrum, we have shown that it is difficult to fit all data better than a null model without these sterile neutrinos.
This difficulty persists even when including additional free parameters in the cosmological model, such as a constant dark energy equation of state, curvature of the universe, running of the spectral index, effective number of neutrinos, and 
primordial helium abundance. Thus, if laboratory experiments continue to favor a scenario with two massive sterile neutrinos, and that is shown to be at odds with cosmological observations, then one may have to look towards a more exotic cosmological model than explored here.

However, we have also shown that the viability of a sterile neutrino model is critically sensitive to our ability to identify and control the systematic uncertainties associated with the datasets included in our analysis.
In particular, the sterile neutrino model fits SN data better than the null model when using the MLCS light curve fitter and worse than the null model when using the SALT2 fitter. These differences between the fitters
can be traced back to different assumptions about the nature of color variations in type Ia SNe and the respective U-band models determined in the training.
In a combined analysis of CMB, Hubble constant, and galaxy power spectrum data, along with SN distance measurements, we find that our sterile neutrino model fits the data equally well as the null model
if we employ the MLCS light curve fitter.
Thus, a minimally extended model with two massive sterile neutrinos could be taken to constitute a realistic cosmological scenario, 
and we advocate caution in interpreting combined analyses of cosmological datasets given their different systematic uncertainties.

\section{Conclusions}

Global short-baseline neutrino oscillation data seem to favor the existence of two sterile neutrinos with masses close to 1~eV each (assuming effectively massless active species).
We have studied the extent to which these two neutrinos are allowed by a combination of probes including the cosmic microwave background, Hubble constant, galaxy power spectrum, and supernova distances. 
In the analysis of SN data, we considered the impact on our results of both the SALT2 and MLCS light curve fitters. In particular, we showed that the choice of the SN light curve fitting method has a major impact on the inferred cosmological model.

We find that the sterile neutrino model provides a good fit to each of the considered datasets, and no single probe manages to decisively disfavor the sterile neutrino model with respect to the null model.
In the joint analysis, sterile neutrinos are allowed by the cosmological data ($\Delta\chi^2_{\rm eff} \approx 0$) when using the MLCS light curve fitter for the SNe in the SDSS compilation, and strongly disfavored by the data ($\Delta\chi^2_{\rm eff} \approx 18$) when using the SALT2 fitter for SNe in the Union2 compilation. 
When excluding the supernova measurements, the sterile neutrinos are disfavored by the other datasets at $\Delta\chi^2_{\rm eff} \approx 12$. 
For a 3+1 sterile neutrino model, it is conceivable that the tension is ameliorated, but this depends on the mass of the single sterile neutrino.
As an illustrative comparison, a cosmological model (without sterile neutrinos) that has $w = -0.8$ is disfavored by WMAP+SPT+$P(k)$+HST (no SN data) at the $\Delta\chi^2_{\rm eff} = 9.4$ level compared to the vanilla model with $w = -1$.

If the SALT2 fitter is indicative of the correct way to interpret SN light curve measurements, then reconciling two light ($\sim \rm eV$) sterile neutrinos (consistent with results from short-baseline neutrino oscillation data) with cosmology may require additional freedom in the cosmological model. However, no single parameter from among nonzero curvature, evolving dark energy, additional relativistic species, running of the spectral index, and primordial helium abundance was able to decrease $\Delta\chi^2_{\rm eff}$ or $\Delta {\rm DIC}$ close to zero. In fact, even for an extended space with all of these additional parameters, the sterile neutrino model is mildly disfavored at $\Delta\chi^2_{\rm eff} \approx 3$ (when using the SALT2 fitter).

The important take-home message, however, is that large shifts in $\Delta\chi^2_{\rm eff}$ ($\sim 20$)  already occur from subtle changes to the way parts of the cosmological datasets are analyzed. 
If SN studies converge toward the MLCS fitter (as opposed to the SALT2 fitter), then two sterile neutrinos with masses close to the eV level are easily allowed by the data. 
Interestingly, even when assuming the existence of two massive sterile neutrinos, we continue to find about  $2\sigma$ preference for an additional massless species.
In addition, in this model with two sterile neutrinos, a much larger matter density would be required (by roughly 40\%), which helps preserve the constraint on 
$\sigma_8 (\Omega_m/0.25)^{0.47}$ near the 0.8-mark, in agreement with galaxy cluster abundance measurements. The analysis presented in this paper shows that it is premature to either rule out the existence of two massive sterile neutrinos or claim this model is cosmologically preferred.

\smallskip
{\it Acknowledgements:} 
We much appreciate useful discussions with John Beacom, Ryan Foley, Jan Hamann, Ryan Keisler, Jostein Kristiansen, Gregory Martinez, and Joseph Smidt.
We acknowledge the use of CAMB and CosmoMC packages~\cite{LCL,Lewis:2002ah}. KNA is partially supported by NSF CAREER Grant No.\ 11-59224. MK and SJ were partly supported by NSF Grant No.\ 0855462. This research was supported in part by the Perimeter Institute of Theoretical Physics during a visit by MK. Research at Perimeter Institute is supported by the Government of Canada through Industry Canada and by the Province of Ontario through the Ministry of Economic Development and Innovation.


\begin{thebibliography}{99}
\frenchspacing

\bibitem{Komatsu:2010fb} 
  E.~Komatsu, {\it et al.}, 
  ApJS {\bf 192}, 18 (2011).

\bibitem{dunkleyact}
  J.~Dunkley, {\it et al.},
  Astrophys.\ J.\  {\bf 739}, 52 (2011).

\bibitem{Keisler:2011aw} 
  R.~Keisler, {\it et al.},
  Astrophys.\ J.\  {\bf 743}, 28 (2011).

\bibitem{Joudaki:2012fx} 
  S.~Joudaki,
  arXiv:1202.0005 [astro-ph.CO].

\bibitem{Hamann:2010bk} 
  J.~Hamann, S.~Hannestad, G.~G.~Raffelt, {\it et al.}, 
  Phys.\ Rev.\ Lett.\  {\bf 105}, 181301 (2010).

\bibitem{Archidiacono:2011gq} 
  M.~Archidiacono, E.~Calabrese and A.~Melchiorri,
  Phys.\ Rev.\ D {\bf 84}, 123008 (2011).

\bibitem{Riess:2011yx} 
  A.~G.~Riess, {\it et al.},
  Astrophys.\ J.\  {\bf 730}, 119 (2011).
  
\bibitem{Smith:2011es} 
  T.~L.~Smith, {\it et al.}, 
  Phys.\ Rev.\ D {\bf 85}, 023001 (2012).

\bibitem{Hou:2011ec} 
  Z.~Hou, {\it et al.}, 
  arXiv:1104.2333 [astro-ph.CO].

\bibitem{Hamann:2007pi} 
  J.~Hamann, S.~Hannestad, G.~G.~Raffelt and Y.~Y.~Y.~Wong,
  JCAP {\bf 0708}, 021 (2007).

\bibitem{Hamann:2011hu} 
  J.~Hamann,
  arXiv:1110.4271 [astro-ph.CO].


\bibitem{GonzalezMorales:2011ty} 
  A.~X.~Gonzalez-Morales, R.~Poltis, B.~D.~Sherwin and L.~Verde,
  arXiv:1106.5052 [astro-ph.CO].

\bibitem{Calabrese:2011hg} 
  E.~Calabrese, D.~Huterer, E.~V.~Linder, A.~Melchiorri and L.~Pagano,
  Phys.\ Rev.\ D {\bf 83}, 123504 (2011).

\bibitem{Smith:2011ab} 
  A.~Smith, {\it et al.},
Phys.\ Rev.\ D {\bf 85}, 123521 (2012).

\bibitem{Hamann:2011ge} 
  J.~Hamann, S.~Hannestad, G.~G.~Raffelt and Y.~Y.~Y.~Wong,
  JCAP {\bf 1109}, 034 (2011).

\bibitem{Giusarma:2011ex} 
  E.~Giusarma, {\it et al.},
  Phys.\ Rev.\ D {\bf 83}, 115023 (2011).

\bibitem{Giusarma:2011zq} 
  E.~Giusarma, M.~Archidiacono, R.~de Putter, A.~Melchiorri and O.~Mena,
 Phys.\ Rev.\ D {\bf 85}, 083522 (2012).

\bibitem{Fischler:2010xz} 
  W.~Fischler and J.~Meyers,
  Phys.\ Rev.\ D {\bf 83}, 063520 (2011).
  

\bibitem{deHolanda:2010am} 
  P.~C.~de Holanda and A.~Y.~.Smirnov,
  Phys.\ Rev.\ D {\bf 83}, 113011 (2011).

\bibitem{Nakayama:2010vs} 
  K.~Nakayama, F.~Takahashi and T.~T.~Yanagida,
  Phys.\ Lett.\ B {\bf 697}, 275 (2011).

\bibitem{Burenin:2012uy} 
  R.~A.~Burenin and A.~A.~Vikhlinin,
  arXiv:1202.2889 [astro-ph.CO].

\bibitem{Ciuffoli:2012yd} 
  E.~Ciuffoli, J.~Evslin and H.~Li,
  JHEP {\bf 1212}, 110 (2012).

\bibitem{Kopp:2011qd} 
  J.~Kopp, M.~Maltoni and T.~Schwetz,
  Phys.\ Rev.\ Lett.\  {\bf 107}, 091801 (2011)
  [arXiv:1103.4570 [hep-ph]].

\bibitem{Giunti:2011gz} 
  C.~Giunti and M.~Laveder,
  Phys.\ Rev.\ D {\bf 84}, 073008 (2011)
  [arXiv:1107.1452 [hep-ph]].

\bibitem{Donini:2012tt} 
  A.~Donini, P.~Hernandez, J.~Lopez-Pavon, M.~Maltoni and T.~Schwetz,
  JHEP {\bf 1207}, 161 (2012).

\bibitem{Abazajian:2012ys} 
  K.~N.~Abazajian, M.~A.~Acero, S.~K.~Agarwalla, A.~A.~Aguilar-Arevalo, C.~H.~Albright, S.~Antusch, C.~A.~Arguelles and A.~B.~Balantekin {\it et al.},
  arXiv:1204.5379 [hep-ph].

\bibitem{Conrad:2012qt} 
  J.~M.~Conrad, C.~M.~Ignarra, G.~Karagiorgi, M.~H.~Shaevitz and J.~Spitz,
  arXiv:1207.4765. 

\bibitem{DiBari:2001ua} 
  P.~Di Bari,
  Phys.\ Rev.\ D {\bf 65}, 043509 (2002)
  [Addendum-ibid.\ D {\bf 67}, 127301 (2003)].

\bibitem{Abazajian:2002bj} 
  K.~N.~Abazajian,
  Astropart.\ Phys.\  {\bf 19}, 303 (2003)
  [astro-ph/0205238].

\bibitem{Abazajian:2004aj} 
  K.~Abazajian, N.~F.~Bell, G.~M.~Fuller and Y.~Y.~Y.~Wong,
  Phys.\ Rev.\ D {\bf 72}, 063004 (2005).

\bibitem{Hannestad:2012ky} 
S.~Hannestad, I.~Tamborra and T.~Tram,
  JCAP {\bf 1207}, 025 (2012)
  [arXiv:1204.5861 [astro-ph.CO]].
  
\bibitem{Mirizzi:2012we} 
  A.~Mirizzi, N.~Saviano, G.~Miele and P.~D.~Serpico,
  arXiv:1206.1046 [hep-ph].

\bibitem{Kneller:2004jz} 
  J.~P.~Kneller and G.~Steigman,
  New J.\ Phys.\  {\bf 6}, 117 (2004).


\bibitem{Simha:2008zj} 
  V.~Simha and G.~Steigman,
  JCAP {\bf 0806}, 016 (2008).

\bibitem{Steigman:2007xt} 
  G.~Steigman,
  Ann.\ Rev.\ Nucl.\ Part.\ Sci.\  {\bf 57}, 463 (2007).
    
  
\bibitem{Peimbert:2007vm} 
  M.~Peimbert, V.~Luridiana and A.~Peimbert,
  Astrophys.\ J.\  {\bf 666}, 636 (2007).
  
\bibitem{Izotov:2007ed} 
  Y.~I.~Izotov, T.~X.~Thuan and G.~Stasinska,
  Astrophys.\ J.\  {\bf 662}, 15 (2007).
  
\bibitem{Izotov:2010ca} 
  Y.~I.~Izotov and T.~X.~Thuan,
  Astrophys.\ J.\  {\bf 710}, L67 (2010).

\bibitem{Aver:2010wq} 
  E.~Aver, K.~A.~Olive and E.~D.~Skillman,
  JCAP {\bf 1005}, 003 (2010).
  
\bibitem{Aver:2010wd} 
  E.~Aver, K.~A.~Olive and E.~D.~Skillman,
  JCAP {\bf 1103}, 043 (2011).

\bibitem{Aver:2011bw} 
  E.~Aver, K.~A.~Olive and E.~D.~Skillman,
  arXiv:1112.3713 [astro-ph.CO].
    

\bibitem{Lewis:2002ah} 
  A.~Lewis and S.~Bridle,
  Phys.\ Rev.\ D {\bf 66}, 103511 (2002).

\bibitem{cosmomclink}
\url{http://cosmologist.info/cosmomc/}

\bibitem{Reid:2009xm} 
  B.~A.~Reid, {\it et al.},
  Mon.\ Not.\ Roy.\ Astron.\ Soc.\  {\bf 404}, 60 (2010).

\bibitem{Amanullah:2010vv} 
  R.~Amanullah, {\it et al.},
  Astrophys.\ J.\  {\bf 716}, 712 (2010).

\bibitem{Kessler:2009ys} 
  R.~Kessler, A.~Becker, D.~Cinabro, J.~Vanderplas, J.~A.~Frieman, J.~Marriner, T.~MDavis and B.~Dilday {\it et al.},
  Astrophys.\ J.\ Suppl.\  {\bf 185}, 32 (2009).

\bibitem{Joudaki:2011nw} 
  S.~Joudaki and M.~Kaplinghat,
  Phys.\ Rev.\ D {\bf 86}, 023526 (2012)
  [arXiv:1106.0299 [astro-ph.CO]].

  \bibitem{Smith}
Smith,~R.~E., {\it et al.}, MNRAS, {\bf 341}, 1311 (2003).

\bibitem{Rudd}
Rudd,~D., Zentner,~A.~R., Kravtsov,~A.V., Astrophys.\ J., {\bf 672}, 19 (2008).

\bibitem{vanDaalen:2011xb} 
  M.~P.~van Daalen, J.~Schaye, C.~M.~Booth and C.~D.~Vecchia,
  Mon.\ Not.\ Roy.\ Astron.\ Soc.\  {\bf 415}, 3649 (2011).

\bibitem{McDonald}
McDonald,~P., Trac,~H., Contaldi,~C., MNRAS, {\bf 366}, 547 (2006).

\bibitem{Joudaki} 
  S.~Joudaki, A.~Cooray and D.~E.~Holz,
  Phys.\ Rev.\ D {\bf 80}, 023003 (2009).

\bibitem{Saito:2008bp}
  S.~Saito, M.~Takada and A.~Taruya,
  Phys.\ Rev.\ Lett.\  {\bf 100}, 191301 (2008).

\bibitem{Brandbyge:2008rv}
  J.~Brandbyge, S.~Hannestad, T.~Haugboelle and B.~Thomsen,
  JCAP {\bf 0808}, 020 (2008).

\bibitem{Wong:2008ws}
 Y.~Y.~Y.~Wong,
 JCAP {\bf 0810}, 035 (2008).

\bibitem{Saito:2009ah}
  S.~Saito, M.~Takada and A.~Taruya,
  Phys.\ Rev.\  D {\bf 80}, 083528 (2009).

\bibitem{Guy:2007dv} 
  J.~Guy, P.~Astier, S.~Baumont, D.~Hardin, R.~Pain, N.~Regnault, S.~Basa and R.~G.~Carlberg {\it et al.},
  Astron.\ Astrophys.\  {\bf 466}, 11 (2007).

\bibitem{Jha:2006fm} 
  S.~Jha, A.~G.~Riess and R.~P.~Kirshner,
  Astrophys.\ J.\  {\bf 659}, 122 (2007).

\bibitem{Guy:2010bc} 
  J.~Guy, M.~Sullivan, A.~Conley, N.~Regnault, P.~Astier, C.~Balland, S.~Basa and R.~G.~Carlberg {\it et al.},
  Astron.\ Astrophys.\  {\bf 523}, A7 (2010).


\bibitem{Foley:2010mm} 
  R.~J.~Foley, A.~V.~Filippenko, R.~Kessler, B.~Bassett, J.~A.~Frieman, P.~M.~Garnavich, S.~W.~Jha and K.~Konishi {\it et al.},
  arXiv:1010.2749 [astro-ph.CO].

\bibitem{rjfoley}
  R.~J.~Foley, private communication.

\bibitem{LCL}
Lewis,~A., Challinor,~A., Lasenby,~A., Astrophys.\ J., {\bf 538}, 473 (2000).

\bibitem{camblink}
\url{http://camb.info}

\bibitem{gelmanrubin}
  A.~Gelman, D.~B.~Rubin, Statist.~Sci., {\bf 7}, 457 (1992).

\bibitem{Cortes:2007ak} 
  M.~Cortes, A.~RLiddle and P.~Mukherjee,
  Phys.\ Rev.\ D {\bf 75}, 083520 (2007).

\bibitem{spiegelhalter}
D.~Spiegelhalter, N.~G.~Best, B.~P.~Carlin, {\it et al.}, 
J.~Royal.~Stat.~Soc.~B 64 583�639 (2002).

\bibitem{Trotta:2008qt} 
  R.~Trotta,
  Contemp.\ Phys.\  {\bf 49}, 71 (2008)
  [arXiv:0803.4089 [astro-ph]].

\bibitem{bridlelewis}
M.~P.~Hobson, {\it et al.},
{\it Bayesian Methods in Cosmology}, University Press, Cambridge, 2010.

\bibitem{Kristiansen:2011mp} 
  J.~R.~Kristiansen and O.~Elgaroy,
  arXiv:1104.0704 [astro-ph.CO].
 
\bibitem{joskri}
  J.~R.~Kristiansen, private communication.

\bibitem{Vikhlinin:2008ym} 
  A.~Vikhlinin, {\it et al.},
  Astrophys.\ J.\  {\bf 692}, 1060 (2009).

\bibitem{Sanchez:2012sg} 
  A.~G.~Sanchez, {\it et al.},
  arXiv:1203.6616 [astro-ph.CO].

\bibitem{Percival:2009xn} 
  W.~J.~Percival {\it et al.}  [SDSS Collaboration],
  Mon.\ Not.\ Roy.\ Astron.\ Soc.\  {\bf 401}, 2148 (2010)

\bibitem{Kaplinghat:2003vf} 
  M.~Kaplinghat and S.~Bridle,
  Phys.\ Rev.\ D {\bf 71}, 123003 (2005)
  [astro-ph/0312430].


  
\end{thebibliography}
\end{document}